# Surveying the Extreme Sky with *EXIST*

**Jonathan E. Grindlay**
*Harvard Smithsonian Center for Astrophysics*
*Cambridge, MA*
*E-mail:* `josh@cfa.harvard.edu`

**Lorenzo Natalucci[1]**
*Istituto Nazionale di Astrofisica, IASF-Roma*
*Via del Fosso del Cavaliere 100,Roma, Italy*
*E-mail:* `lorenzo.natalucci@iasf-roma.inaf.it`

**on behalf of the *EXIST* Team**

The recent hard X-ray surveys performed by INTEGRAL and *Swift* have started to reveal the demographics of compact sources including Super-Massive Black Holes hosted in AGNs and have proven invaluable in tracking explosive events as the death of massive stars revealed by Gamma-Ray Bursts up to cosmological distances. Whereas the observations have contributed significantly to our understanding of the sources populations in the Local Universe, it has also become evident that revealing the processes that drive the birth and evolution of the first massive stars and galaxies would have required a further big step in both sensitivity and capability to study transient phenomena since their very beginning and covering different wavebands simultaneously. Therefore, after its decennial history as a proposed hard X-ray survey mission, *EXIST* has now turned into a new, more advanced concept with three instruments on board covering the IR/optical and X-ray/soft gamma-ray bands. The *EXIST* new design (Grindlay 2009a) is therefore much improved in its capability for prompt study of GRBs (with autonomous determination of the redshift for many of them) and broadband spectral studies of SMBHs and transients in the high energy band from 0.1 to several hundred keV, with sensitive optical/NIR and soft X-ray identifications and followup studies.

## 1.Introduction

The *EXIST* concept based on a sensitive survey hard X-ray observatory has been proposed several times, starting in the mid 90's (Grindlay et al. 1995). It was one of the three recommended missions in the 2001 Decadal Survey, together with GLAST (now launched) and Con-X (that evolved into the IXO concept as a joint program of NASA, ESA and JAXA). *EXIST* was proposed to the most recent Decadal Survey (Astro2010) with submissions to the two Requests for Information (RFI-1 and -2) in April and August 2009. After the recent results of *Swift* especially in the study of Gamma-Ray Bursts (GRB), the mission profile has much evolved from the concept of a *pure* hard X-ray mission. Beyond the High Energy Telescope (HET), which is the main instrument with its unprecedented large area and coded mask imaging capability, two complementary payload instruments have been implemented in the design: the optical-Infrared Telescope (IRT) and the Soft X-ray Imager (SXI) (see Figure 1). The HET is a wide field (~90° x 70°) hard X-ray coded mask imager with a very large area (4.5m$^2$) CdZnTe (CZT) detection plane, featuring sub-mm resolution and low power ASIC readout. The IRT is an optical/infrared telescope with passively cooled mirrors giving exceptional sensitivities in the NIR band (AB=24 in 100s). Its 1.1m telescope is already flight proven since a very similar (optical) telescope is now flying on an Earth observing (*Geo-Eye1*) satellite. Finally the SXI, contributed by Italy/ASI, is a soft X-ray Wolter type-I telescope coupled to an X-ray detector camera with CCD detector, sensitive in the 0.1-10 keV range, with an effective area similar to that of one XMM/*Newton* module.

---

[1] Speaker

  *EXIST* is proposed for launch in mid 2017 in a Low Earth Orbit at lower inclination than *Swift,* aboard an EELV carrier (e.g. AtlasV-401, with 4m fairing). The combined use of the three instruments on board *EXIST* will allow study with unprecedented sensitivity of the most extreme high energy sources in the Universe, the high-z GRBs, which will be re-pointed promptly by the Spacecraft by autonomous trigger based on the <20" hard X-ray localization on board. Some ~600 GRBs/yr are expected, including ~7-10% at high redshift (z >7). Most of the GRBs followed up, including the farthest ones (up to z~20 if they exist) will have their redshift autonomously measured by IRT following the detection of the Lyman-alpha break in the spectra of their afterglows. Given the fast re-pointing (~150s), SXI will be able to measure images and soft X-ray (0.1-10 keV) spectra from the early afterglow and provide independent determination of the redshift through the detection of X-ray absorption features. The IRT and SXI together will be able to provide information about the physical properties of the circum-burst environment, resulting in a systematic study of the properties of the galaxies hosting the GRBs. Apart from ~2 orbits of followup on each of ~2GRBs per day, for the first 2 years of the mission *EXIST*/HET will continuously scan the entire sky every 2 orbits detecting about 40,000 AGN. For the next 3years of pointed followup studies of AGN and transients, an additional ~20,000 AGN will be detected and located by the wide-field HET.

  The *EXIST* concept, in the present form, exploits fully the scientific heritage of the *Swift* and INTEGRAL missions as well as their instrument teams expertise. The *EXIST* spacecraft bus will be built around the model developed for *Fermi*. The payload configuration of *EXIST* has minimal development risks as the underlying technology for the three instruments is already developed and flight proven.

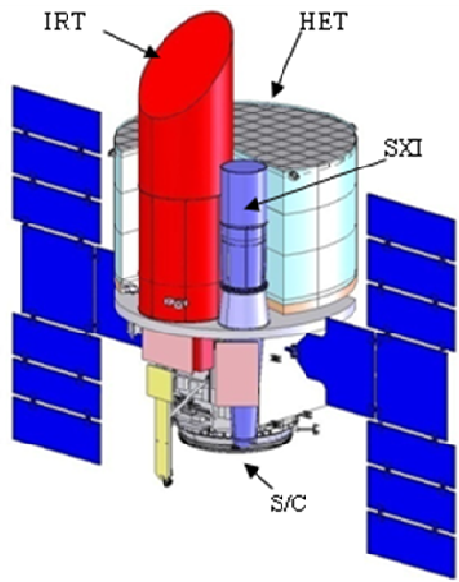

**Figure 1.** 3-D view of the *EXIST* spacecraft and payload. Its primary instrument is the High Energy Telescope (HET), a wide field coded aperture instrument covering the 5-600 keV energy band and imaging sources in a ~90° x 70° field of view with ≤20" positions. Also shown is the Soft X-ray Imager (SXI), 0.1-10 keV with an effective area of 950cm$^2$ at 1.5 keV and 3.5m focal length, and the IRT, an optical-IR 1.1m aperture telescope, passively cooled to -30C, and covering the 0.3 – 2.2 micron range with both imaging and spectroscopy. The IRT pixel size is 0.15 arcsec and its Field-of-View in Imaging mode is 16 arcmin$^2$.

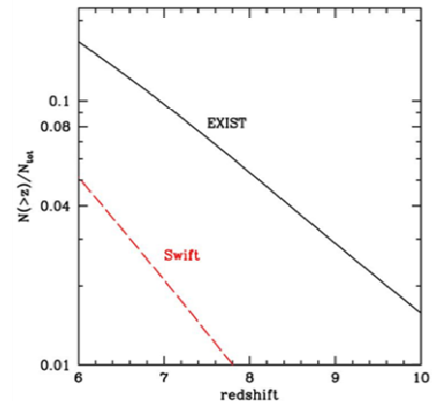

**Figure 2** Integral distribution of the fractional number of bursts expected to be detected by *EXIST*, compared to those obtained by *Swift*. Computations based on Salvaterra (2008).

This paper is organized as follows. In section 2 a highlight of the topics within the main science objectives of *EXIST*. Then the operational scenario is described (Section 3) and a description of the instruments is given in Section 4.

## 2. Science objectives

The science objectives for the new *EXIST* mission have been extensively investigated during the recently funded NASA/ASMC study. They have been streamlined along three main directions: a) the study of the GRBs in order to map the first stars and structure (Epoch of Reionization) in the early Universe; b) the study of SuperMassive Black Holes (SMBH) to reveal their nature, masses, evolution and integrated luminosities; and c) the study of X-ray transients of all types, and to probe the nature of compact objects and their progenitors.

### 2.1 Gamma-ray Bursts

The discovery of the connection of the subclass of the Long Soft Bursts (LSB), which connect the majority of GRBs with the death of massive stars mostly occurring in star forming galaxies, opens new opportunities for *EXIST* on the study of the early Universe and in particular, the Epoch of Re-ionization (EOR) in which the first massive stars and galaxies are believed to be formed and re-ionized the Universe. GRBs are the most appropriate probes of the EOR, thanks to their high luminosity and also to their relatively simple and smooth power law spectra. High-z GRBs are known to exist since *Swift* has detected 3 at z >6.4, including the most recent record at z=8.2 (Tanvir et al 2009). Prompt spectra of their afterglow emission can reveal important information on the re-ionization structure and global history. The IRT sensitivity to the afterglow emission of GRBs is shown in Figure 3 against the flux of a well known GRB. *EXIST* will be able to confirm, for example, the existence of the very massive, metal-poor *PopIII* stars which could have affected the environment of the early Universe, and, in general, put constraints on potential sources of re-ionization. As shown by recent simulations (Grindlay 2009a), the *EXIST*/IRT will be able to measure the damping wing of the Lyα absorption features in GRB afterglows which McQuinn et al (2008) have shown to be sensitive to the IGM neutral fraction and to the size of the HII bubbles which are expected to enshroud the GRB hosts. Moreover, the observation of this feature together with the SXI measurements of the soft X-ray absorption can potentially put constraints on the fraction of escaping ionizing photons (McQuinn 2009).

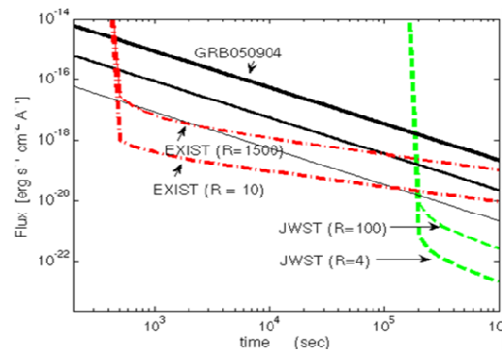

**Figure 3** Limiting flux, vs time of the IRT spectral measurements (for resolutions R = 10 vs. 1500) against the flux of a GRB with a flux of 1, 0.1, and 0.01 the value of GRB050904.

### 2.2 SuperMassive Black Holes and Surveys

The results of surveys performed by *Swift* and INTEGRAL, with the detection of ~1000 hard X-ray sources have allowed to discover and study in more detail new classes of high energy sources, mostly coincident with accretion dominated AGNs and Galactic Binaries. Highly obscured objects, not easily detected below 10 keV are now known to be numerous both in AGN and within Galactic systems. *EXIST* will be able to determine unambiguously the fraction of Compton Thick (CT) AGN sources and constrain their contribution to the cosmic background by observation of a few thousands of them up to a redshift z~1 (see Della Ceca et al. 2010 for this and other AGN discoveries likely for *EXIST*). An example of the capabilities of *EXIST* for spectral studies of CT AGN is given in Figure 4. It must be also emphasized, that the spectral coverage capability (0.1-600 keV) provided by HET and SXI is unique to disentangle and monitor unambiguously the direct and reprocessed radiation in the different states.

In the field of jet-powered sources, blazars have been confirmed to be powerful emitters in hard X-rays and their high energy spectrum may reveal important information about their state and Spectral Energy Density (SED) evolution. The recent studies of blazars have shown the need for simultaneous coverage of the spectra and variability in different wavebands, due to the large extension in wavelength of the involved processes and the high level of variability.

Multiwavelength observations of blazars at relatively low redshift, like those recently performed in the X-ray, gamma-ray, optical/IR and radio have shown themselves to be a powerful tool to both characterize the SED and determine the relative variability of the two main peaks giving insight to the geometry and physics of the emission region (e.g., Costamante 2009). In particular, X-ray/soft γ–ray spectra can be either dominated by the synchrotron or Compton peak and thus of great help in the monitoring of the spectral evolution of the SED. *EXIST* will be able to observe many thousands of these objects simultaneously in the near IR/optical and X-ray/soft-gamma ray bands and to determine the redshift for most of them. Moreover, blazars are probably the best sources for the investigation of the formation and evolution of SMBH in the early Universe, and the high sensitivity of *EXIST*, coupled to its multiband coverage will allow to observe and identify the most luminous objects at least up to z~8 (if they exist) with ~25 objects expected at z>6 (Della Ceca et al. 2010). Currently, only a handful of the *Fermi* blazars (Abdo et al. 2009) consisting of BL Lacs and Flat Spectrum Radio Quasars have been detected in hard X-rays by INTEGRAL (Ubertini et al. 2009) and *Swift* (Ajello et al. 2009) up to a redshift z~4. *EXIST* can complement and greatly extend *Fermi* during the all sky survey period (see Section 3). The large *EXIST* blazar sample will provide unique information on the SED of the powerful, highest-z (>4) blazars not accessible to *Fermi* and provide the best constraints on blazars as probes of the first SMBHs out to z ~8 (Ghisellini et al 2009).

**2.3 The Transient Universe**

*EXIST* will explore a new phase space of transient phenomena in the Universe, by the study of a wide variety of transient events: first, the extragalactic emission associated with short hard bursts (SHBs) and their direct association with gravitational wave detections from Advanced LIGO (Bloom et al 2009). Then for long GRBs, prompt locations and optical-NIR identifications from *EXIST* will enable identification of sub-luminous GRBs with Supernovae (SN-Ic) explosions. Within the Local Group, *EXIST* will reveal outbursts of stellar mass compact objects, and flares from highly magnetized Soft γ–ray Repeaters. And dormant SMBHs will be revealed out to ~200Mpc in the act of tidally disrupting a star (Grindlay 2004, Gezari 2009). Additional goals for study of transients are related to: a) exploiting techniques to pinpoint SNe at the moment of explosion, as discovered for SN2008D by Soderberg et al (2009), by detection of the X-ray pulse emitted by the shockwave ripping through the surface of the star (~10 events/year) ; b) investigating more deeply the mechanisms for magnetic field generation in stars; c) study and characterization of outbursts from stellar mass and intermediate-mass Black Holes (IMBH) and the first census of isolated BH and IMBH systems accreting from molecular clouds to determine their numbers and spatial distribution (Grindlay 2009b); and d) exploring a new phase space of the transient Universe in parallel with complementary studies at longer wavelengths (synergy with LSST, ALMA, SKA, and LISA). Other interesting subjects are related to the studies of sub-luminous GRB (see below) and to the Supergiant Fast X-ray Transients, recently discovered in our Galaxy and identified as a new

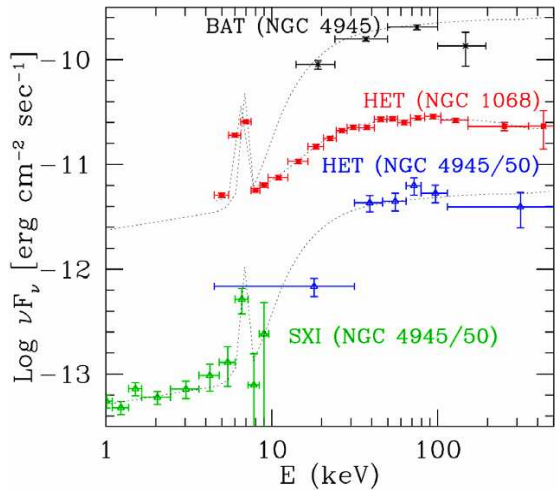

**Figure 4**. Simulation (Della Ceca et al 2009) of *EXIST* SXI+HET observations of well known Compton Thick AGNs (NGC 4945 and NGC1068). NGC 4945 has a BAT measured flux of ~3x10$^{-10}$ erg cm$^{-2}$ s$^{-1}$ (15-200 keV, shown in black). A heavily absorbed source with same spectrum but flux 50X fainter can be studied and its spectrum resolved into components. HET spectra are simulated survey spectra whereas SXI data are for a 10$^4$s followup pointing.

class by the INTEGRAL survey and *Swift* follow-up studies (Sidoli et al. 2010 and refs. therein).

Subluminous GRBs belong to the Long GRB class and are thus associated with the explosions of massive stars. They occur at low redshifts (z<0.2) and their luminosity is lower than high redshift GRBs (Soderberg et al. 2006). Four of these events have been associated to Type Ibc SNae (see e.g. Gehrels et al. 2006). The relationship between the emission of a GRB produced by the collapse of a massive star and the possible occurrence of a SN event is not clear, nor it is the nature of the progenitors of this class. The current GRB sample is largely incomplete, but the large increase in sensitivity of *EXIST* over *Swift*/BAT and its lower threshold would increase significantly (by factors of >10) the number of objects and shed light on their origin and on the properties of their hosts. Interestingly, INTEGRAL/IBIS (which is more sensitive than Swift to GRB detection but has a much smaller Field-Of-View and no fast followup capability) has recently detected a population of faint, sub-luminous GRBs with long lags and spatial distribution consistent with a nearby location of the hosts (Ubertini et al. 2009b).

## 3. *EXIST* Operations Summary

During the first two years *EXIST* will be programmed for a full sky survey of HET, whereas a second phase (3 years, assuming a nominal mission lifetime of 5 years) is foreseen for the follow-up of sources discovered during the survey. The survey will be performed by the HET conducting an orbital scan at the zenith with an offset of ±30 degrees (towards the north and the south on alternate orbits, respectively) for all-sky survey coverage each 3h. GRB follow-up will be performed during both the scanning and pointed mission phases. GRBs and transient source positions are calculated on board within ~10s with a <20" error radius and the autonomous re-pointing takes ~150s. A Guest Investigator (GI) program will support the entire mission: for the first two years it will be based on the *Fermi* type with proposals for individual targets and classes of targets, whereas for the survey follow-up it will be modeled on the *Swift* and *Chandra* GI programs.

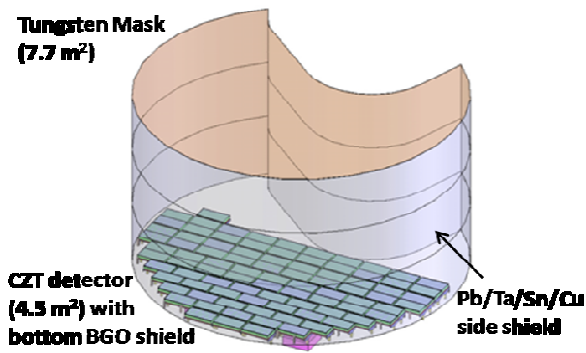
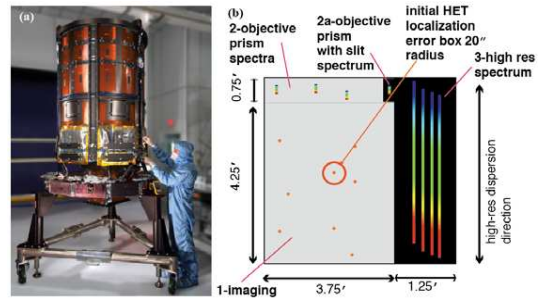

**Figure 5**. Schematic of the HET on board EXIST. The position of the 88 modules composing the CZT detection plane is shown. Each module includes a BGO bottom shield.

**Figure 6**. Left: the *Geo-Eye* Telescope, very similar to the IRT. Right: the allocation of the IRT FOV composed by: imaging (1), low resolution (R=30) spectroscopy (2 & "A) and high resolution (R=3000) spectroscopy (3).

## 4. The *EXIST* Payload

### 4.1 High Energy Telescope (HET)

The High Energy Telescope on board *EXIST* is a wide FoV imager, sensitive in the 5-600 keV band that employs large arrays of fine pixel CZT detectors and a hybrid (double spatial scale) tungsten mask (see Figure 5). The design of the detection plane is based on a hierarchy of components: 88 identical modules, each with 128 Detector Crystal Units (DCUs). Each DCU is made of a single CZT crystal subdivided in 32x32 pixels of 0.6mm pitch. The 1024 channels of each DCU are readout by a specific ASIC, which is a further evolution of the low power ASIC developed for *NuSTAR*. The hybrid mask will be built using 15mm pixels etched into a laminate of 10 x 0.3 mm thick tungsten sheets and surrounding a single 0.3mm thick tungsten sheet with etched 1.25mm open pixels. The combined laminate gives a 25% open fraction below 100 keV and 50% above 200 keV. This will provide efficient shielding against diffuse background and allow for a rapid coarse mode analysis for transients localization. The HET 1year scanning survey full-sky sensitivity (5σ) is expected to be ~0.08-1.5 mCrab, depending on the energy range. See Hong et al. (2010) for a more complete description of the HET.

### 4.2 Optical IR Telescope (IRT)

The IRT Optical Telescope Assembly (OTA) is based on an optical design similar to the *Geo-Eye* 1.1m telescope built by ITT (see Figure 6), but with simpler Cassegrain configuration. Its focal plane enables both imaging and spectroscopic capabilities to cover a broad spectral range, from 0.3-2-2µm in four bands centered at 0.41, 0.71, 1.14 and 1.71 µm. Each of the two NIR bands has a dedicated readout system using a 2K x 2K H2RG array (similar to that developed for NIRSPEC on JWST). Both broadband photometry and spectroscopy of the same field will be possible. The two visible detectors are *HyViSi* detectors with readout ASICs similar to the H2RG channels. All four detectors are 2K×2K with 0.15" pixel size. The FoV for each of the three primary modes is shown in Figure 6 (right panel). The high-resolution spectral channel covers ~25% of the detector and includes 4" spatial coverage along the slit (~25 pixels). Imaging and slit-less spectra are read out for survey coverage, even during targeted slit spectra acquisition. The IRT will be operational only during pointing mode. Its 5σ sensitivity in Imaging mode is AB=24 in 100s in all four bands. For spectra, the sensitivities are AB ~23 (for resolution R =30) and AB ~20 (R =3000) for 3000s integrations.

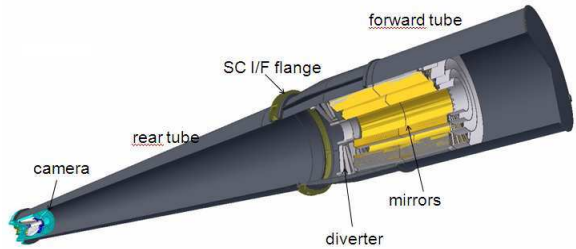

**Figure 7**. View of the SXI with details of parts (Mirrors, camera, structure)

### 4.3 Soft X-ray Imager (SXI)

The proposed design of the SXI (see Figure 7) is based on a Wolter type-I mirror assembly with 26 nested shells and a focal plane camera with pnCCD detector. The focal plane distance is 3.5m and the max. diameter of the mirrors is 60cm. The effective area of the instrument is ~950cm$^2$ at 1.5 keV and greater than 100cm$^2$ at 8 keV. The angular resolution of the mirror system (Half Power Diameter) is estimated to be less than 20 arcsec throughout the whole FoV (Tagliaferri et al. 2009). The FoV of SXI is 30'x30' and its sensitivity is 2x10$^{-15}$ cgs for an exposure time of 10$^4$s. A notable characteristics of the SXI is its ability to perform a sensitive survey of half the sky during the zenith scanning (Natalucci et al. 2009) portion of the mission. The characteristics of the camera design are very similar to those of the XRT on *Swift* and EPIC on *XMM*. In order to operate efficiently during the survey it will have a frame readout speed between 5 and 10 ms. The limiting sensitivity of the survey is ~5x10$^{-14}$ erg cm$^{-2}$ s$^{-1}$ in two years. Since during the scanning the IRT does not operate, SXI will be able to improve the localization of many faint HET sources from ~20" to about 1-2".

**Acknowledgements.** JG and the US-based *EXIST* team acknowledge support from NASA grant NNX08AK84G, and LN acknowledges the support of ASI by grant I/088/06/0.